\documentclass[conference]{IEEEtran}
\usepackage{amsmath,amsfonts}
\usepackage{algorithmic}
\usepackage{algorithm}
\usepackage{array}
\usepackage{textcomp}
\usepackage{stfloats}
\usepackage{url}
\usepackage{verbatim}
\usepackage{graphicx}
\usepackage{cite}
\usepackage{subcaption} 
\usepackage{bm}
\begin{document}

\title{RowHammer Vulnerability Counter (RVC): Redefining RowHammer Detection with Victim-Centric Tracking}

 \author{Lavi Jain, Venkata Kalyan Tavva

\thanks{Lavi Jain and Venkata Kalyan Tavva are with IIT Ropar, Rupnagar, Punjab (e-mail: {2023csm1005, kalyantv }@iitrpr.ac.in). }
}

\markboth{IEEE Journal of \LaTeX\ Class,~Vol.~12, No.~6, February~2024}%
{Shell \MakeLowercase{\textit{et al.}}: A Sample Article Using IEEEtran.cls for IEEE Journals}


\maketitle
\begin{abstract}
The Rowhammer vulnerability poses an increasing challenge with newer generations of DRAM and aggressive technology scaling. Existing mitigation techniques, such as Graphene, Twice, and Hydra, primarily rely on tracking activation counts for each row and issuing refreshes when a row reaches a predefined tracking threshold. However, these methods have inherent limitations, including inefficiencies in identifying rows genuinely at risk of bit flips.  

In this paper, we propose a novel framework called \textbf{Rowhammer Vulnerability Count (RVC)}, which shifts the focus from activation count tracking to evaluating a row's actual vulnerability to bit flips. By selectively issuing refreshes only to rows on the verge of experiencing bit flips, RVC drastically reduces unnecessary refresh operations. We also demonstrate that prior works have incorrectly set tracking thresholds, leading to security flaws.  

Our evaluation shows that RVC achieves \textbf{95–99.99\% improvement} in mitigation induced refreshes when compared to Graphene, with no additional space overhead. Furthermore, RVC improves energy efficiency and reduces average LLC latency by up to \textbf{76.91\%}, making it a highly efficient and scalable solution for addressing Rowhammer in modern DRAM systems. These findings establish RVC as a superior approach for preventing Rowhammer, outperforming existing methods in both accuracy and efficiency.
\end{abstract}


\section{Introduction}
\label{intro}
DRAM is ubiquitous technology for main memory design. RowHammer is a well-known vulnerability in modern DRAM devices, wherein frequent accesses (or ``hammering") to a row (``aggressor row") can induce bit-flips in adjacent rows (``victim rows"). The maximum number of accesses an aggressor row can sustain without inducing bit-flips in victim rows is known as the Rowhammer Threshold($T_{rh}$) of the device. This vulnerability, may lead to data corruption through unintended bit flips, posing a significant threat to data integrity and security. Due to technology scaling, RowHammer thresholds have decreased from 140K~\cite{rowhammer} to about 4.8K~\cite{revisitingrowhammer}, and is expected to reach lower thresholds in the coming years. To prevent these induced bit-flips several counter-based RowHammer detection techniques have been proposed like \cite{graphene},\cite{hydra},\cite{start},\cite{rowhammercache}, etc. These techniques primarily focus on monitoring the activation counts of rows and initiate mitigation as and when a row reaches a predefined threshold (``Tracking Threshold"). One of the mitigation approaches is issuing a Victim Row Refresh (VRR) command to refresh the physically adjacent victim rows of an aggressor row. However, this method suffers from significant drawbacks. 
The first drawback is for any aggressor based approach, the second drawback if the victims are only accessed recently then it is could be drawback for any aggressor based approach. The second example is for accessing victims rows.

Drawbacks:- (i) The first problem with keeping track of activation count of just the aggressor rows is the assumption that high row activation count directly correlates with increased victim risk. In reality, this correlation may be incorrect because the victim row(s) may have been accessed within the current observation window (typically refresh interval, $t_{REFI}$). (ii) Second, if the victim row is common among multiple aggressor rows, it could have been recently refreshed due to some other aggressor, hence resetting their vulnerability to Rowhammer induced bit-flips. In such a case, tracking individual aggressors and initiating VRR based on them will result in unnecessary refreshes. 

For example:-
\begin{figure}[!t]
\centering
\includegraphics[width=3.5in]{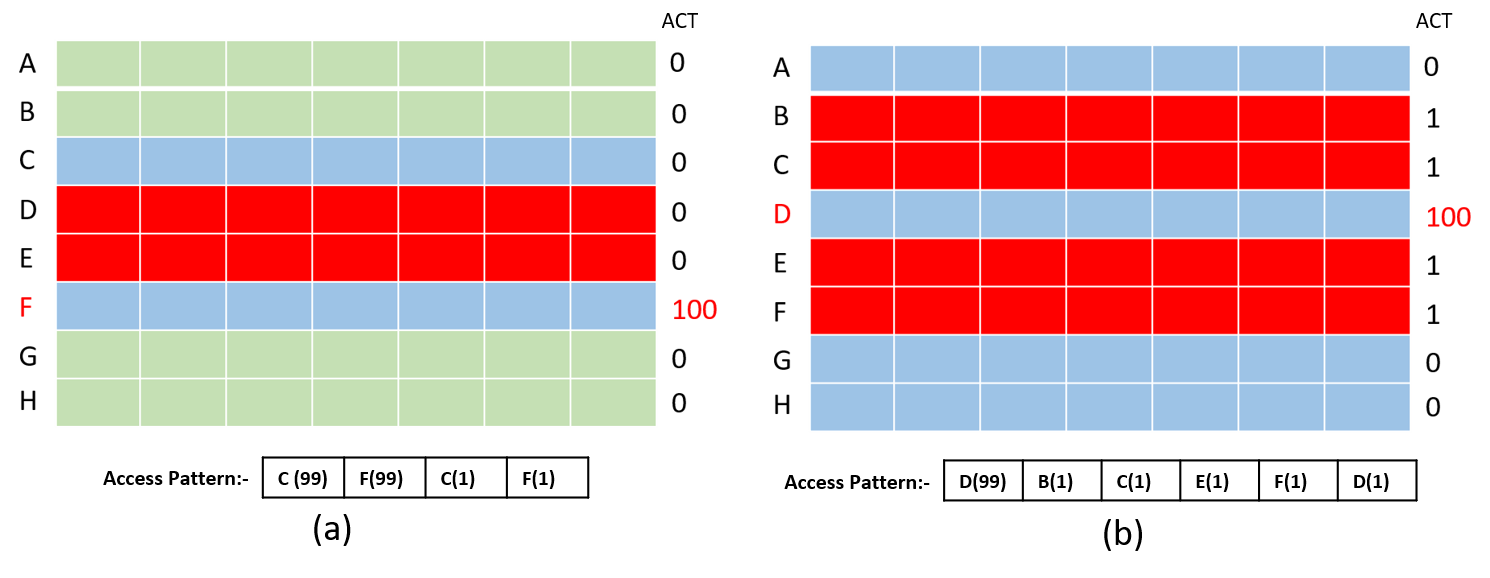}
\caption{Problems in Aggressor Activation Count Tracking with VRR Rowhammer Mitigation: (a) Highlights the issue of refreshing rows that were recently refreshed, leading to redundant refreshes (red-marked rows). (b) Illustrates the problem of refreshing rows that were recently accessed.} 
\label{fig_1_1}
\end{figure}
\begin{itemize}
\item In case of Figure~\ref{fig_1_1}a, we see that although rows A, B, D, E were refreshed due to row C reaching the Threshold, we redundantly refresh rows D, E again  when row F reaches thresholds.
\item In Figure~\ref{fig_1_1}b, rows B, C, E, and F were accessed just a few activations before row D reached its threshold. Since accessing a row inherently refreshes the row by restoring its charge, refreshing these rows again when row D reaches the threshold is redundant and unnecessary .
\end{itemize}

\begin{figure}[!t]
\centering
\includegraphics[width=2in]{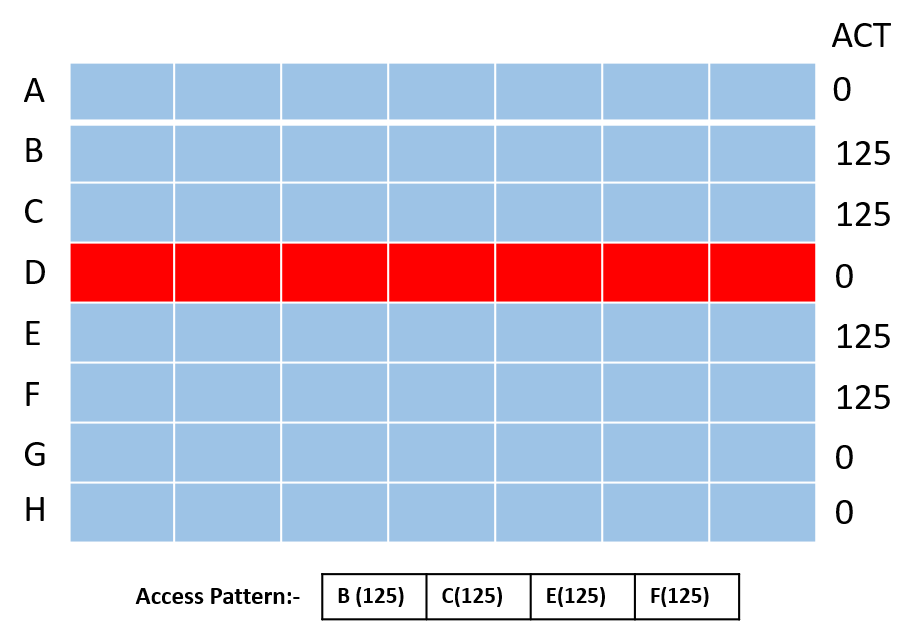}
\caption{Cumulative effect on a victim by accessing adjacent rows on a system with $T_{rh}$ of 500 and a blast radius of 2.}
\label{fig_2}
\end{figure}

Third drawback with existing tracking methods is when defining the Tracking Threshold (T) for a given Rowhammer Threshold ($T_{rh}$). To address the problem of cumulative effect of multiple aggressor rows on a shared victim row (Figure~\ref{fig_2}), $T_{rh}$ must be reduced by a factor of $2n$, where $n$ represents the number of rows affected by an aggressor row in one direction, commonly referred to as the ``blast radius". This reduction accounts for the cumulative stress on a victim row resulting from multiple aggressor rows in its vicinity. While this is conservative, it is defined to ensure robustness against RowHammer attacks, particularly common victim attacks. Such a design choice inadvertently impacts benign workloads leading to a significant performance drop. In typical scenarios, benign applications may access only a limited number of rows near the victim, and no RowHammer attack may be occurring. By universally reducing $T_{rh}$ by $2n$, assuming the worst case attack scenario, leads to frequent trigger of mitigation.

To address these drawbacks, we demonstrate that tracking victim rows instead of aggressor rows provides a more accurate and power-efficient approach in mitigating RowHammer, while also improving performance with reduced memory access latency. 
 In this work, we propose the Rowhammer Vulnerability Counter (RVC) that provides several advantages over naive aggressor activation count based row tracking methods:
\begin{enumerate}
    \item Reduction in number of mitigations triggered and refreshes: On average, RVC achieves a 92.1\% reduction in VRR refreshes and 91.9\% fewer mitigation commands issued.
    \item Robust Protection: RVC provides effective defense against single-sided, double-sided, and multi-sided RowHammer attacks by accurately tracking victim vulnerability.
    \item By reducing excessive refreshes, RVC decreases total DRAM energy consumption by an average of 8.5\% and VRR-induced energy consumption by over 91.6\%. Furthermore, by alleviating DRAM contention, it improves average LLC latency by up to 28.7\%, leading to a more efficient memory system. 
\end{enumerate}
Through detailed evaluation, we demonstrate that our modified version of Graphene~\cite{graphene} adopted with RVC, outperforms the original aggressor activation count-based tracking mechanism in Graphene. By focusing on victims RVC provides a secure and efficient approach to detect and mitigate rowhammer vulnerability in current DRAM systems.

\section{Background}
\label{bckgnd}
A RowHammer bit flip can occur at a DRAM row if the neighboring rows have been activated frequently and the number of activations within a refresh interval has crossed the RowHammer Threshold (Trh). The degree to which a DRAM device is susceptible to RowHammer attack is characterized
 by Trh. This number has decreased rapidly with newer
 generations of DRAM technology. In the latest DDR-5 chips,
 the threshold is speculated to be around 500 whereas a decade
 back TRH used to be 140K\cite{rowhammer}. It is interesting to note that
 even benign applications can frequently surpass such ultra-low
 Trh thresholds. A RowHammer attacker can corrupt any data
 including data corresponding to page tables [46] located at any unspecified DRAM locations. A RowHammer attack can also be induced remotely [17]. So, a RowHammer tracker should be agile and should track the row activations and trigger the mitigation, well ahead of time so that rowhammer induced bit flips can be avoided.
Rowhammer Tracking :-
The tracking mechanism works by keeping track of ACTs to the rows. It issues mitigation, once the counter of any row reach a predefined threshold. Such rows are called aggressor rows, while the nearby rows susceptible to bit-flips are called victim rows. The number of victim rows on each side of the aggressor is called as the blast-radius. 
\subsection{Various RowHammer Trackers}
Various kinds of RowHammer trackers have been proposed to mitigate these attacks, categorized as follows:
\begin{itemize}
    \item SRAM-based Trackers: Utilizing SRAM to store activation counters for tracking row activations.
    \item DRAM-based Trackers: Embedding tracking mechanisms directly within DRAM to monitor activations.
    \item Hybrid Trackers: Combining SRAM and DRAM techniques to leverage the advantages of both.
    \item LLC-based Trackers: Using the last-level cache (LLC) to monitor and manage row activations.
\end{itemize}

\subsection{Determining Threshold (T)}
To prevent Rowhammer-induced bit-flips, the tracking threshold (\(T\)) is set to be significantly lower than the Rowhammer threshold (\(T_{\text{rh}}\)) of the system. This precaution is necessary due to the following two scenarios:
\subsubsection{Activations across Refresh Windows}
The maximum number of activations (\(ACT\)) that a row can recieve without triggering a victim row refresh occurs across two consecutive refresh windows. Specifically, a row can receive up to \(T-1\) activations in one refresh window and another \(T-1\) activations in the next, without refreshing the victim rows of the aggressor row. Therefore, to account for this, the maximum number of activations a row can experience without refreshing its victims is calculated as \(2(T-1)\).  

\subsubsection{Effect of Activation on a common victim}
Previous studies have shown that activations to a single row can induce bit-flips in up to \(2n\) victim rows, where \(n\) represents the blast radius above and below the aggressor row. To securely prevent such bit-flips, the tracker must initiate mitigation whenever a row's activation count reaches \(\frac{T_{\text{rh}}}{2n}\).  
An illustrative example is provided in Fig.2, where 125 activations to rows B, C, E, and F in a single refresh window lead to bit-flips in row D. This demonstrates the importance of accurately accounting for the cumulative disturbance effect on a common victim row.

So finally the effective formula is 
\begin{equation}
2(T-1) < \frac{T_{rh}}{2n}
\end{equation}
This formula has been highlighted previously in Graphene~\cite{graphene}. Determining the correct tracking threshold is essential for the prevention of Rowhammer induced bit-flips. Techniques such as Hydra, RowhammerCache, Comet, Mint, and others have proposed innovative tracking strategies. However, these approaches fail to account for the cumulative effect of activations on a common victim row when defining the tracking threshold. Similarly, DSAC-TRR does not consider activations occurring across multiple refresh windows. These limitations reduce the reliability of their analyses and leave their mechanisms vulnerable to Rowhammer-induced bit-flips, highlighting the importance of carefully setting the tracking threshold to enhance resilience against any such RowHammer attack.
\subsection{Tracker Entries with Misra-Gries Algorithm}
In the context of Rowhammer, the Misra-Gries algorithm guarantees that any row address activated more than \(\frac{W}{N_{\text{entry}} + 1}\) times during the last \(W\) activations is included in the count table. To ensure tracking of rows activated more than T times in the last W activations, the number of entries must satisfy
\begin{equation}
N_{\text{entry}} > \frac{W}{T} - 1
\end{equation}
\section{RowHammer Vulnerability Counter}
To improve the security and efficiency of existing "aggressor" counter based rowhammer detection approaches, we propose the Rowhammer Vulnerability Counter(RVC), that effectively protects against single, double, and multi-sided adversarial access patterns, while also reducing unnecessary mitigations in the system.
The main idea behind RVC is to monitor the cumulative effect on a "victim row" caused by accesses to its neighboring rows, effectively tracking how "vulnerable" a row is to experiencing a bit-flip. By adopting this approach, the mechanism aims to selectively refresh only those rows that are "at risk" of undergoing a bit-flip, thereby optimizing refresh operations and reducing overhead.
\subsection{Working of RVC}
\subsubsection*{Initialization/Every Refresh Interval}
\begin{enumerate}
\item{Reset the counter of each row to 0.}
\end{enumerate}
\subsubsection*{ Access to Row}
\begin{enumerate}
\item{Reset its count to 0.}
\item{Increment the count of above and below rows((\( \pm \))blast radius) by 1}
\item{If any row's count reaches T, we issue refresh for that row only.}
\end{enumerate}
When a row is accessed, RVC resets its counter to 0, as accessing the row is same as refreshing it. Simultaneously, the counters of rows within a predefined blast radius (e.g., two rows above and two rows below the accessed row) are incremented by 1, accounting for the increased risk of disturbance from nearby activations. The system operates with a threshold value, T, such that if any row’s counter exceeds T, it triggers a refresh on that row to prevent potential bit flips.
\subsection{Determining the Threshold T}
To determine the Threshold(T) we build up on equation 1, where division by 2n accounts for the effect of activations on a common victim across the blast radius as shown in Fig2 in a conventional setup. In the novel approach described in this work, the counter of rows +-blast radius is incremented directly. As a result, the cumulative activation tracking inherently captures all affected rows, making the division by 2n redundant. Consequently, the equation is simplified to: \begin{equation}
2(T-1) < T_{rh}
\end{equation}
. where 
T is the threshold for our RVC tracker approach.
.This adjustment ensures that for a given row the cumulative effect on it due to activation's encompassing all adjacent rows, remains below Trh, thereby accurately detecting rows on the verge of row hammer-induced bit-flips.
\subsection{Selective VRR Command}
With each activation (ACT), the counters of rows within the blast radius are incremented. When a row's count reaches the threshold \(T\), it is selectively refreshed. However, multiple rows within the blast radius may reach the threshold simultaneously. To handle this efficiently, we propose a new Selective VRR (Victim Row Refresh) command. 

The Selective VRR command uses a bit-vector of size \(2 \times \text{blast radius}\) bits, which is sent to the DRAM,along with the address of the row whose activation caused row's to reach the threshold T in the tracker. The DRAM internally decodes and then performs selective refresh operations only for the rows indicated by the bit-vector near the specified address. This approach minimizes unnecessary refreshes and optimizes system performance by refreshing only the rows at risk of Rowhammer-induced bit-flips.

\subsection{RVC Tracker Entries with Misra-Gries Algorithm}
In the RVC framework, each activation increases the count for rows within the blast radius, effectively in equation (2) scaling the total number of activations \(W\) and the threshold \(T\) by a factor of \(2n\)(as shown in Equation (3)). This proportional scaling ensures that the tracker size remains constant, requiring no additional area overhead when using Misra-Gries algorithm for tracking victims.
\subsection{Example}
For example, consider a DRAM configuration with a RVC threshold T = 500. If a row, A, is accessed 495 times, followed by one access each to the two rows above (A-1, A-2) and below (A+1, A+2), and finally 5 more accesses to row A, the system would not trigger a refresh. This outcome is because the counters of A-1,A-2,A+1,A+2 are reset to 0 when directly accessed. So even if row A is accessed 500 times, there are 0 refreshes that needs to be triggered in such a scenario.

By incrementing counters for adjacent rows and resetting the accessed row’s counter, the system efficiently protects against single, double and multi-sided access patterns. The selective refresh mechanism minimizes overhead, preventing excessive refresh operations while ensuring robust protection against rowhammer-induced bit flips.
This method provides a improvement in performance and security by keeping count for victims responding to memory access patterns, making it an enhancement over purely aggressor-based counter designs.
\section{Evaluation}
We used Ramulator2 to perform simulations for our evaluation, with the experimental setup illustrated in Table 1(Should i add voltage details for energy computation?).

\begin{table}[!t]
\caption{Simulation Setup\label{tab:sim_setup}}
\centering
\begin{tabular}{|c|c|c|}
\hline
\multicolumn{3}{|c|}{\textbf{Core}} \\
\hline
\textbf{Parameter} & \textbf{1 Core} & \textbf{4 Core} \\
\hline
Core Type & \multicolumn{2}{c|}{Out-of-Order (OoO), 3.2Ghz} \\
\hline
Last Level Cache (LLC) & 2MB & 8MB (2MB/core, Shared) \\
\hline
LLC MSHRs per Core & 16 & 16 per core \\
\hline
LLC Associativity & 8 WAY & 8 WAY \\
\hline
\multicolumn{3}{|c|}{\textbf{Memory Controller}} \\
\hline
Address Mapping & \multicolumn{2}{c|}{RoBaRaCoCh} \\
\hline
Memory Scheduler & \multicolumn{2}{c|}{FRFCFS} \\
\hline
Row Policy & \multicolumn{2}{c|}{ClosedRowPolicy (cap: 4)} \\
\hline
Refresh Manager & \multicolumn{2}{c|}{AllBank} \\
\hline
\multicolumn{3}{|c|}{\textbf{DRAM}} \\
\hline
Type & \multicolumn{2}{c|}{DDR5-16Gb-x8} \\
\hline
Memory Size & \multicolumn{2}{c|}{4GB} \\
\hline
Memory Bus Speed & \multicolumn{2}{c|}{1.6 GHz (3.2 GHz DDR)} \\
\hline
Channels x Ranks x Banks & \multicolumn{2}{c|}{1 x 2 x 4 (8 Bank Groups)} \\
\hline
tRCD - tRP - tCAS & \multicolumn{2}{c|}{16.25 - 16.25 - 16.25 ns} \\
\hline
tRC - tRFC & \multicolumn{2}{c|}{48.125 ns - 295 ns} \\
\hline
Voltage & \multicolumn{2}{c|}{Default DDR5} \\
\hline
\end{tabular}
\end{table}

\begin{table}[ht]
  \centering
  \tiny
  \caption{WorkLoad Mixes}
  \label{tab:benchmarks}
  \begin{tabular}{|l|c|c|c|c|}
    \hline
    Label & Benchmark 1 & Benchmark 2 & Benchmark 3 & Benchmark 4 \\
    \hline
    H0 & 437.leslie3d & 450.soplex   & 470.lbm        & 510.parest \\
    H1 & 437.leslie3d & 510.parest   & 505.mcf        & 459.GemsFDTD \\
    H2 & 470.lbm      & 459.GemsFDTD & 505.mcf        & 437.leslie3d \\
    H3 & 459.GemsFDTD & 470.lbm      & 462.libquantum & 433.milc \\
    H4 & 459.GemsFDTD & 470.lbm      & 549.fotonik3d  & 519.lbm \\
    H5 & 450.soplex   & 462.libquantum & 434.zeusmp    & 538.imagick \\
    H6 & 505.mcf      & 482.sphinx3  & 434.zeusmp     & 523.xalancbmk \\
    H7 & 471.omnetpp  & 505.mcf      & 519.lbm        & 458.sjeng \\
    H8 & 433.milc     & 505.mcf      & 462.libquantum & 523.xalancbmk \\
    H9 & 450.soplex   & 549.fotonik3d  & 437.leslie3d  & 531.deepsjeng \\
    M0 & 434.zeusmp   & 433.milc     & 445.gobmk      & 538.imagick \\
    M1 & 510.parest   & 434.zeusmp   & 435.gromacs    & 544.nab \\
    M2 & 429.mcf      & 482.sphinx3  & 557.xz         & 458.sjeng \\
    M3 & 483.xalancbmk & 470.lbm     & 525.x264       & 526.blender \\
    M4 & 434.zeusmp   & 519.lbm      & 541.leela      & 447.dealII \\
    L0 & 470.lbm     & 456.hmmer    & 435.gromacs    & 523.xalancbmk \\
    L1 & 433.milc    & 508.namd     & 436.cactusADM  & 502.gcc \\
    L2 & 450.soplex  & 403.gcc      & 481.wrf        & 511.povray \\
    L3 & 470.lbm     & 508.namd     & 531.deepsjeng  & 473.astar \\
    L4 & 437.leslie3d & 473.astar   & 435.gromacs    & 525.x264 \\
    L5 & 481.wrf     & 445.gobmk    & 511.povray     & 473.astar \\
    L6 & 557.xz     & 538.imagick  & 456.hmmer      & 520.omnetpp \\
    L7 & 456.hmmer   & 502.gcc      & 436.cactusADM  & 523.xalancbmk \\
    L8 & 447.dealII  & 435.gromacs  & 436.cactusADM  & 526.blender \\
    L9 & 500.perlbench & 436.cactusADM & 544.nab      & 403.gcc \\
    \hline
  \end{tabular}
\end{table}
We evaluate our proposed approach in comparison with the state-of-the-art tracker, Graphene. To facilitate this comparison, we modify the existing Graphene tracker to monitor the \textit{vulnerability count} of victim rows instead of aggressor rows, referring to this modified approach as RVC. For Graphene, we configure the tracking threshold (\(T\)) and counter table size according to Equation 1 and Equation 2. In the case of RVC, we set \(T\) and counter table size as specified in Equation 4 and detailed in Section C.  
For mitigation, Graphene employs the standard VRR command to trigger refreshes when a row’s count reaches the threshold \(T\). In contrast, RVC uses Selective VRR command for mitigation under similar conditions, selectively targeting rows based on their vulnerability count.  

Our evaluation begins with a detailed analysis of System Threshold (Trh) = 500 and Blast Radius = 2, across various heterogeneous 4-core workload mixes outlined in Table 2. Following this, we examine the percentage improvement of our proposed approach over Graphene by varying the blast radius (\(B\)) across values of 1, 2, 4, and 8. The simulation runs for 250M instructions for each core.
We evaluate the effectiveness of RVC through five key metrics:  

\begin{enumerate}  
    \item Mitigation Commands Issued: This metric quantifies the number of mitigations triggered by the tracker.  
    \item Refreshes: We measure the number of row refreshes issued in the DRAM due to RowHammer mitigations. 
    \item Total Energy: This metric captures the overall energy consumptions in DRAM.  
    \item VRR Energy: We measure the specific energy associated with the VRR commands.  
    \item Average LLC Latency: This metric measures the average latency incurred from the time a request is issued to the time it is received back at the LLC. 
\end{enumerate}  

\subsection{Trh 500 and Blast Radius 2}
On a system configured with a threshold (\(Trh\)) of 500 and a blast radius (\(B\)) of 2, our proposed RVC approach demonstrates a significant 92.3\% avg reduction in the number of refreshes triggered by VRR, as illustrated in Figure 4. This substantial improvement highlights the effectiveness of the RVC tracker in minimizing unnecessary refresh operations.  

Additionally, we observe a 91.3\% avg reduction in the number of mitigations commands issued, shown in Figure 3. This improvement can be attributed to two primary factors. First, the increased tracking threshold employed in RVC reduces the frequency of triggering mitigation mechanisms. Second, the RVC tracker efficiently resets the counters for rows accessed, ensuring that only rows identified as highly vulnerable and at imminent risk of bit-flips are refreshed.  

This drastic reduction in both mitigations issued and refreshes directly translates to a significant 91.3\% avg decrease in VRR-induced energy consumption, as depicted in Figure 5. Consequently, this improvement contributes to an 4\% avg reduction in Total DRAM energy consumption, as shown in Figure 6.  

Furthermore, the reduction in refresh operations alleviates contention at the DRAM, allowing them to handle DRAM requests more effectively. This leads to a 5.5\% improvement in average LLC latency, as illustrated in Figure 7.  

These results collectively demonstrate that RVC significantly outperforms Graphene by reducing mitigation overhead, improving energy efficiency, and enhancing memory system performance.  

\subsection{Varying Blast Radius}
In this section, we compare the performance of RVC with Graphene across varying blast radii, specifically for values of 1, 2, 4, and 8.
For higher thresholds and larger blast radii, RVC achieves exceptional gains. For instance, at $T_{\text{rh}} = 5000$ with a blast radius of 2, RVC sets the tracking threshold at 2500 compared to Graphene's 1250. This difference minimizes the number of rows that require refreshes, leading to over 98\% improvement in mitigation's issued. As the blast radius grows, the percentage improvement becomes even more pronounced, reaching up to 99.99\% in case of blast radius 8. 
Additionally, RVC leverages selective refreshes and row count resets on access, ensuring that only critical vulnerable rows are targeted. This leads to drastic reductions in refresh overhead, which directly benefits system performance by allowing DRAM to focus on servicing requests more efficiently. The reduction in refresh operations significantly improves average LLC latency, especially at lower thresholds, where improvements exceed 76.9\% at a blast radius of 8.
\begin{itemize}
    \item VRR Energy: RVC along with Selective VRR achieves consistent energy savings across all thresholds. For example, at $T_{\text{rh}} = 100$, energy improvement ranges from 56.18\% (blast radius 1) to 95.74\% (blast radius 8). At $T_{\text{rh}} = 5000$, savings exceed 96.55\%, peaking at 99.98\%.
    \item Refreshes: RVC reduces refresh operations dramatically. At $T_{\text{rh}} = 100$, the reduction ranges from 57.44\% (blast radius 1) to 96.79\% (blast radius 8). For $T_{\text{rh}} = 5000$, refresh reductions reach up to 99.99\%.
    \item Mitigation Commands issued: Similarly, RVC minimizes victim row refresh (VRR) commands issued. Improvements start at 56.18\% (blast radius 1) and grow to 95.74\% (blast radius 8) at $T_{\text{rh}} = 100$, with higher thresholds achieving nearly complete elimination of redundant commands issued.
    \item Average LLC Latency: By reducing mitigations issued and refreshes, RVC enables DRAM to serve requests faster. At $T_{\text{rh}} = 100$, latency improvement reaches 76.92\% at a blast radius of 8. Even at higher thresholds, RVC outpaces Graphene, ensuring better performance.
\end{itemize}

The RVC Advantage:
\begin{itemize}
    \item Superior Efficiency: By maintaining a consistent threshold and targeting only critical rows, RVC eliminates unnecessary refresh operations, providing unparalleled energy and performance gains.
    \item Scalability with Blast Radius: As the blast radius increases, RVC's advantages become even more pronounced, demonstrating its ability to handle future DRAM devices with lower Rowhammer thresholds with minimal performance degradation.
    \item Enhanced Latency Reduction: RVC's efficient refresh management with Selective VRR ensures that DRAM refreshes are optimized, significantly reducing average latency at the LLC and enhancing overall system responsiveness.
\end{itemize}
Overall, RVC outshines Graphene across all evaluated dimensions. Its ability to deliver consistent improvements in energy efficiency, refresh reduction, and latency underlines its superiority as a rowhammer detection strategy.

\begin{figure}[htbp]
    \centering
    \begin{subfigure}{0.45\textwidth}
        \centering
        \includegraphics[width=\textwidth]{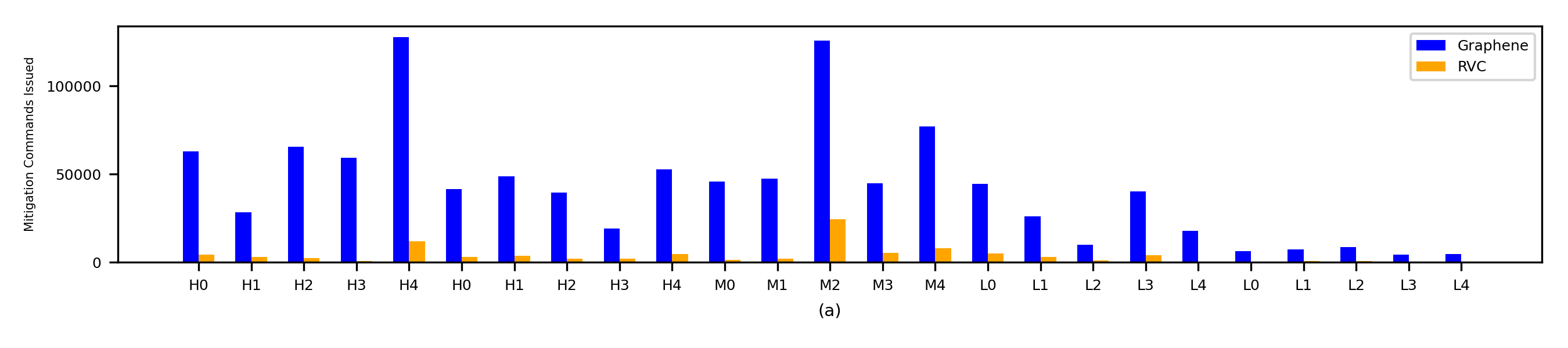}
        \label{fig:plot_a}
    \end{subfigure}
    \begin{subfigure}{0.45\textwidth}
        \centering
        \includegraphics[width=\textwidth]{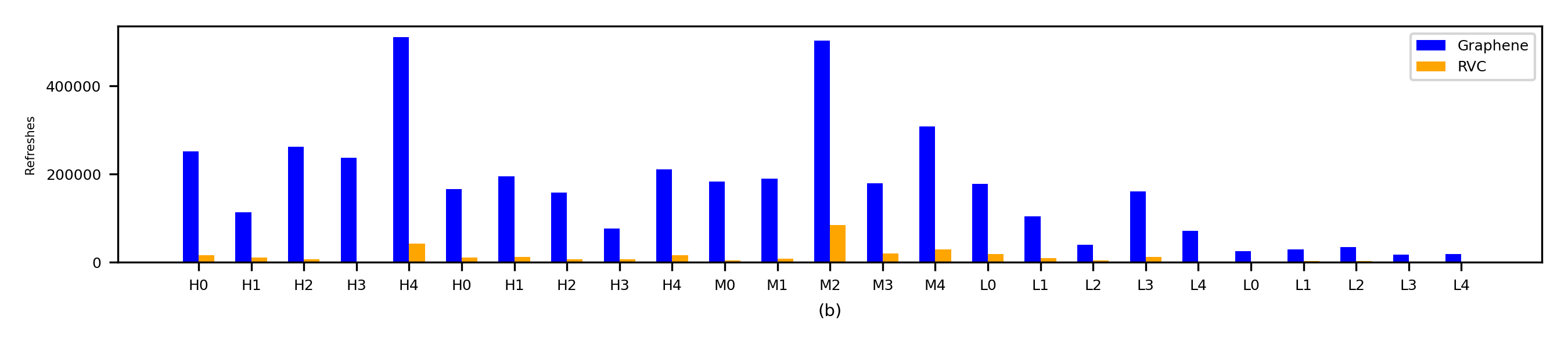}
        \label{fig:plot_b}
    \end{subfigure}
    \begin{subfigure}{0.45\textwidth}
        \centering
        \includegraphics[width=\textwidth]{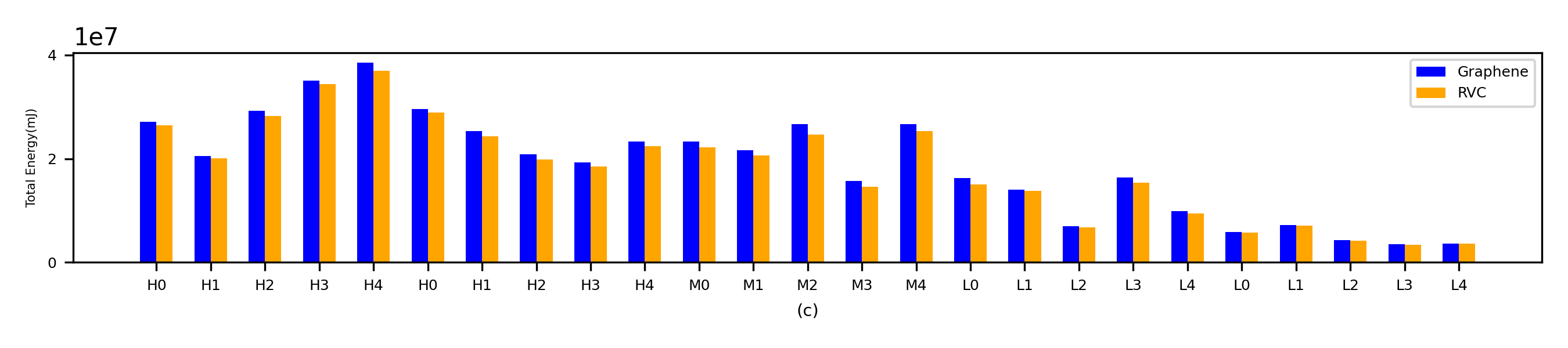}
        \label{fig:plot_c}
    \end{subfigure}
    \begin{subfigure}{0.45\textwidth}
        \centering
        \includegraphics[width=\textwidth]{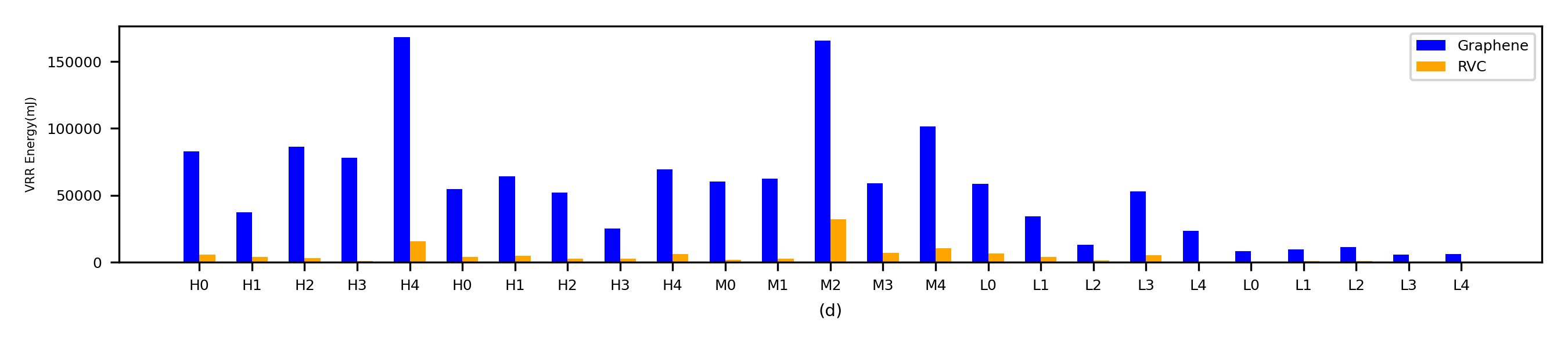}
        \label{fig:plot_d}
    \end{subfigure}
    \begin{subfigure}{0.45\textwidth}
        \centering
        \includegraphics[width=\textwidth]{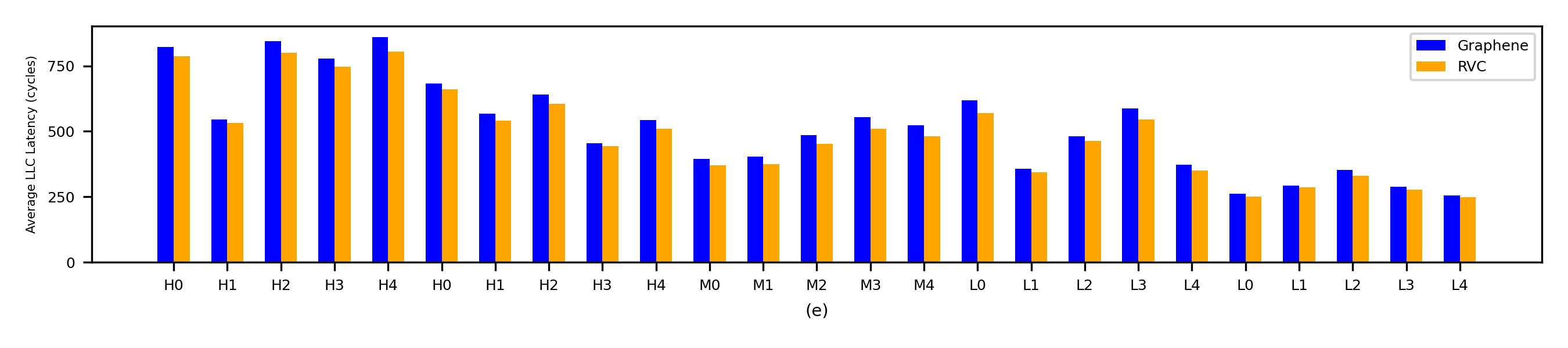}
        \label{fig:plot_e}
    \end{subfigure}

    \caption{Comparison of Graphene and RVC for various metrics.}
    \label{fig:all_plots}
\end{figure}


\begin{figure}[!t]
\centering
\includegraphics[width=3.5in]{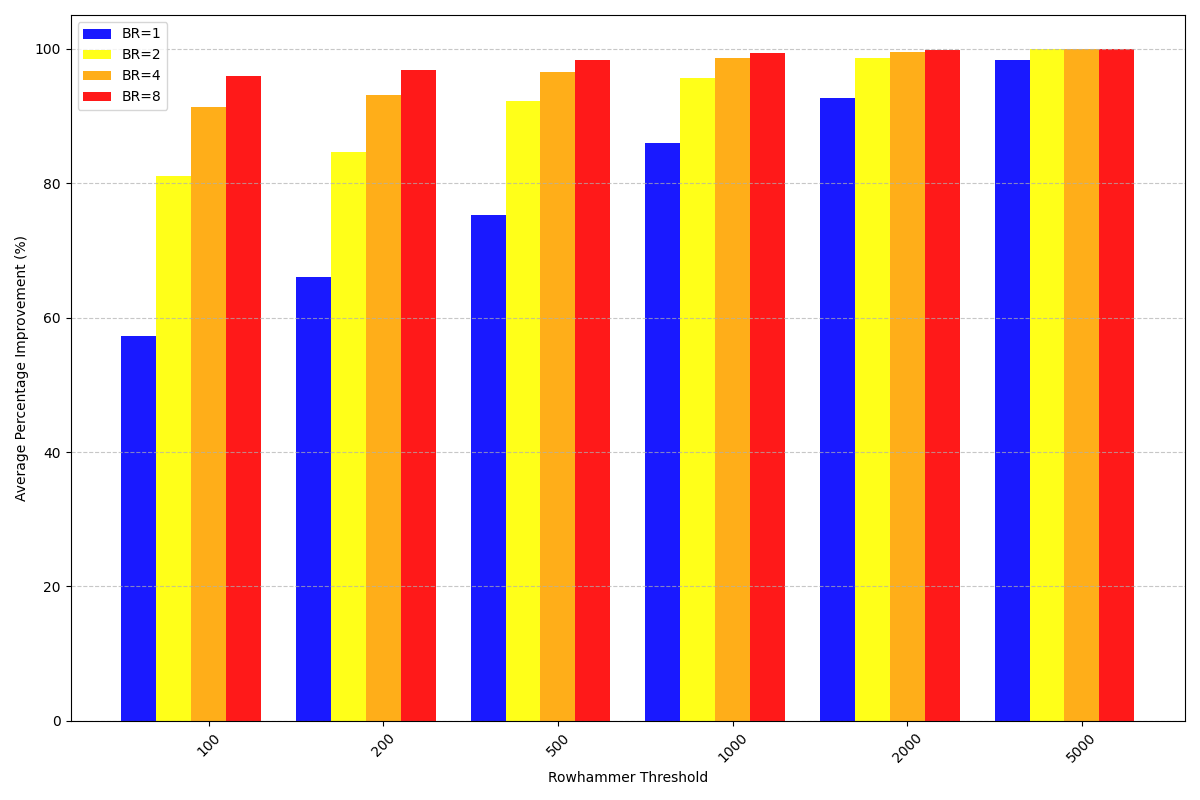}
\caption{Reduction in VRR commands issued.}
\label{fig_vrr_cmd}
\end{figure}
\begin{figure}[!t]
\centering
\includegraphics[width=3.5in]{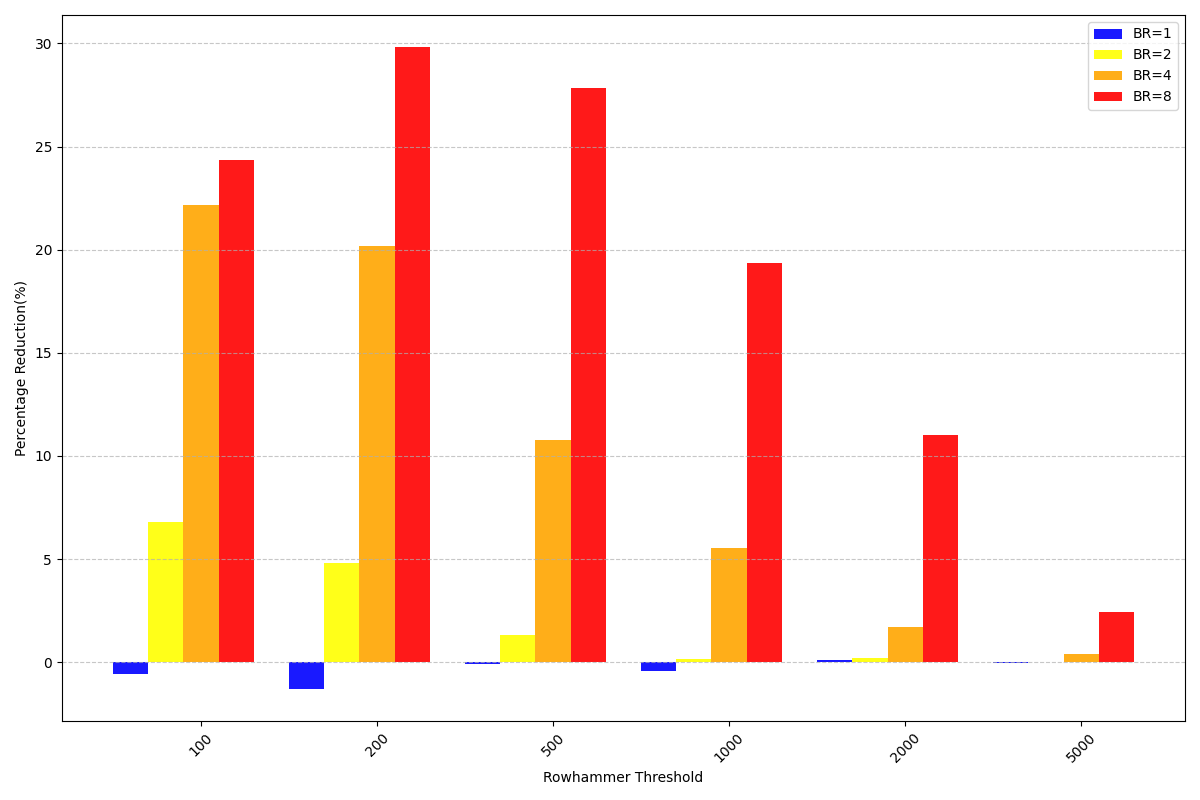}
\caption{Decrease in IPC ANTT.}
\label{fig_ipc}
\end{figure}
\begin{figure}[!t]
\centering
\includegraphics[width=3.5in]{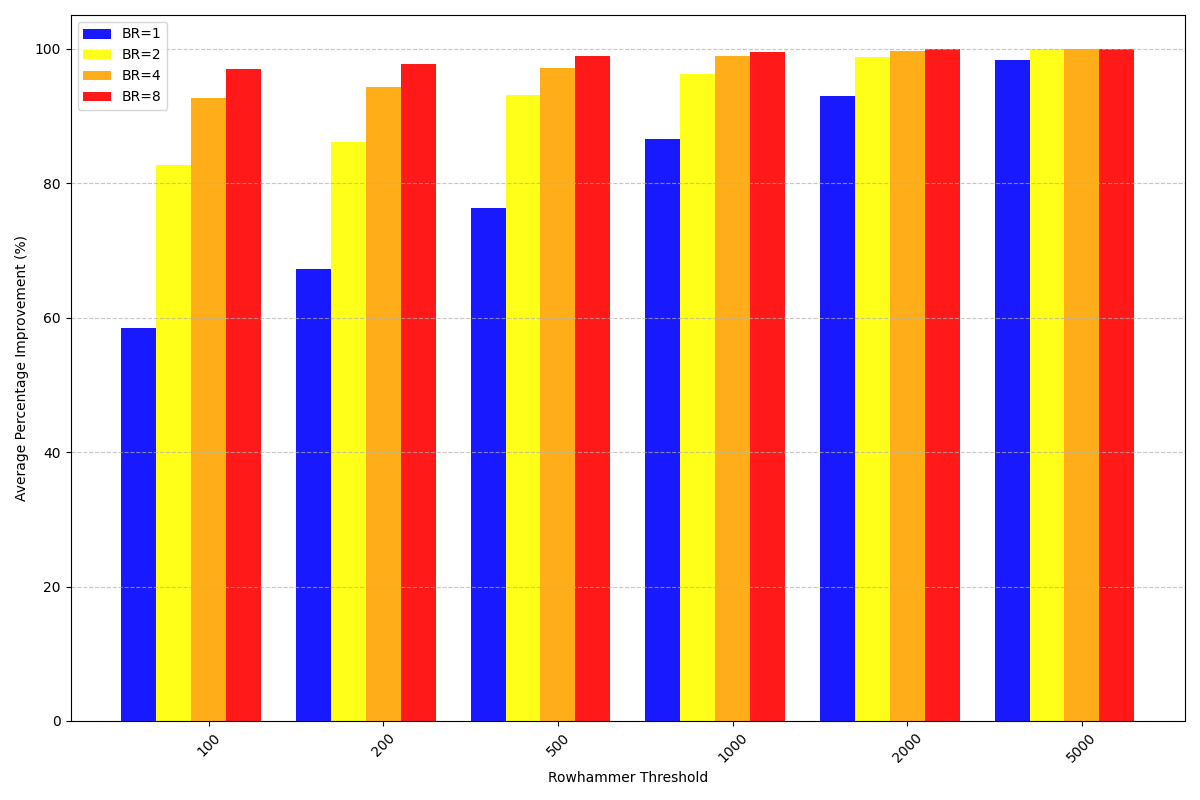}
\caption{Percentage reduction in refreshes.}
\label{fig_ref}
\end{figure}
\begin{figure}[!t]
\centering
\includegraphics[width=3.5in]{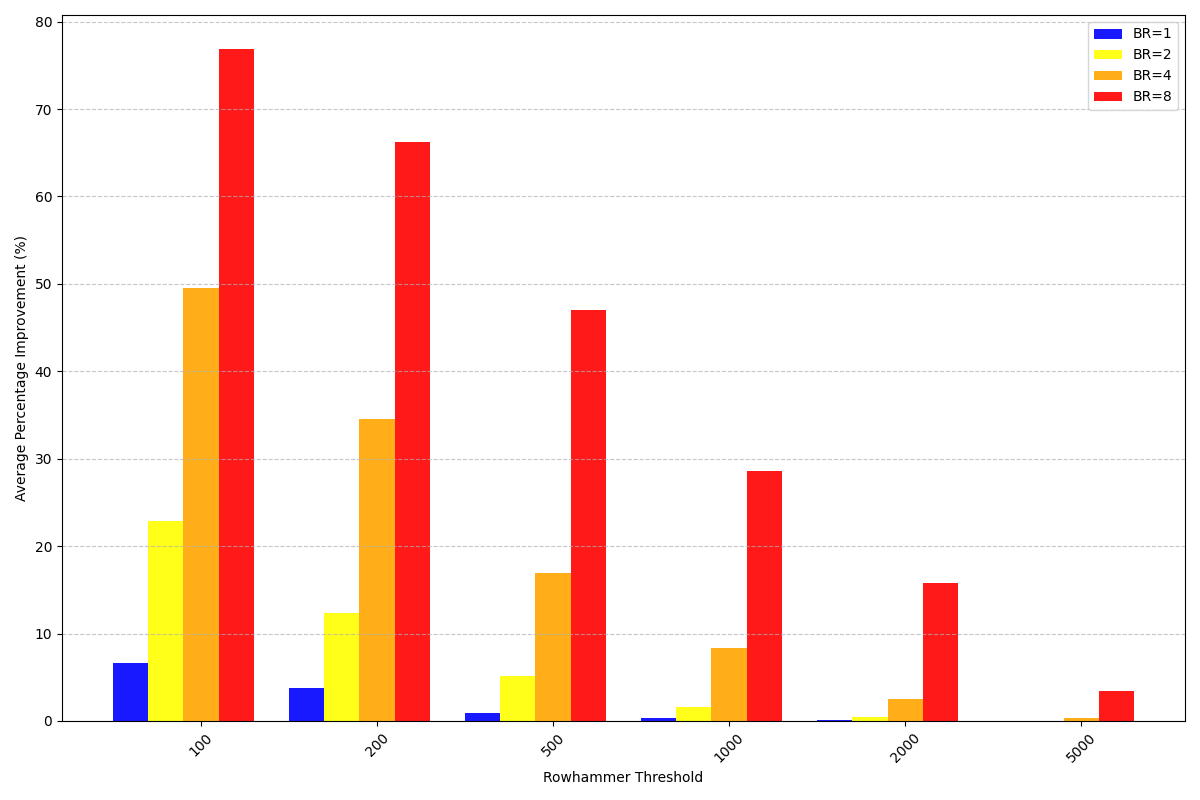}
\caption{Reduction in Latency observed at LLC.}
\label{fig_llc_lat}
\end{figure}
\begin{figure}[!t]
\centering
\includegraphics[width=3.5in]{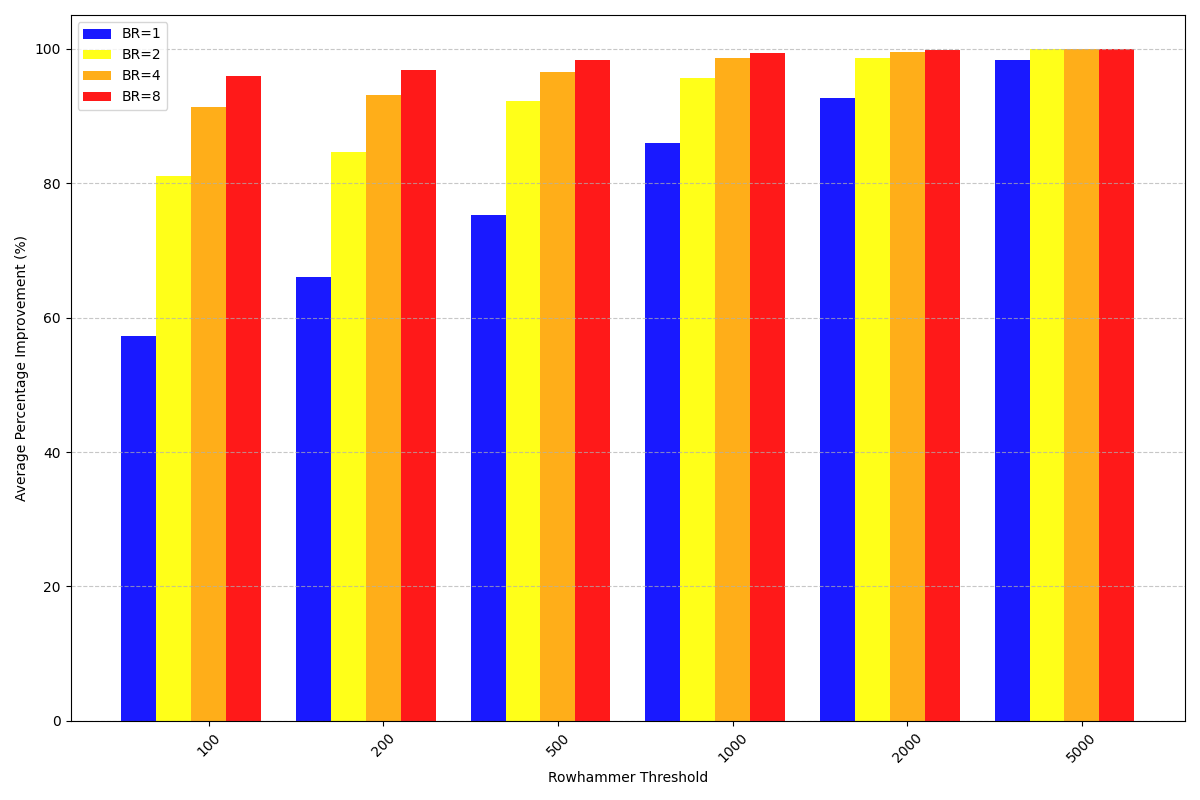}
\caption{Reduction in energy consumed due to VRR.}
\label{fig_energy}
\end{figure}
\begin{figure}[!t]
\centering
\includegraphics[width=3.5in]{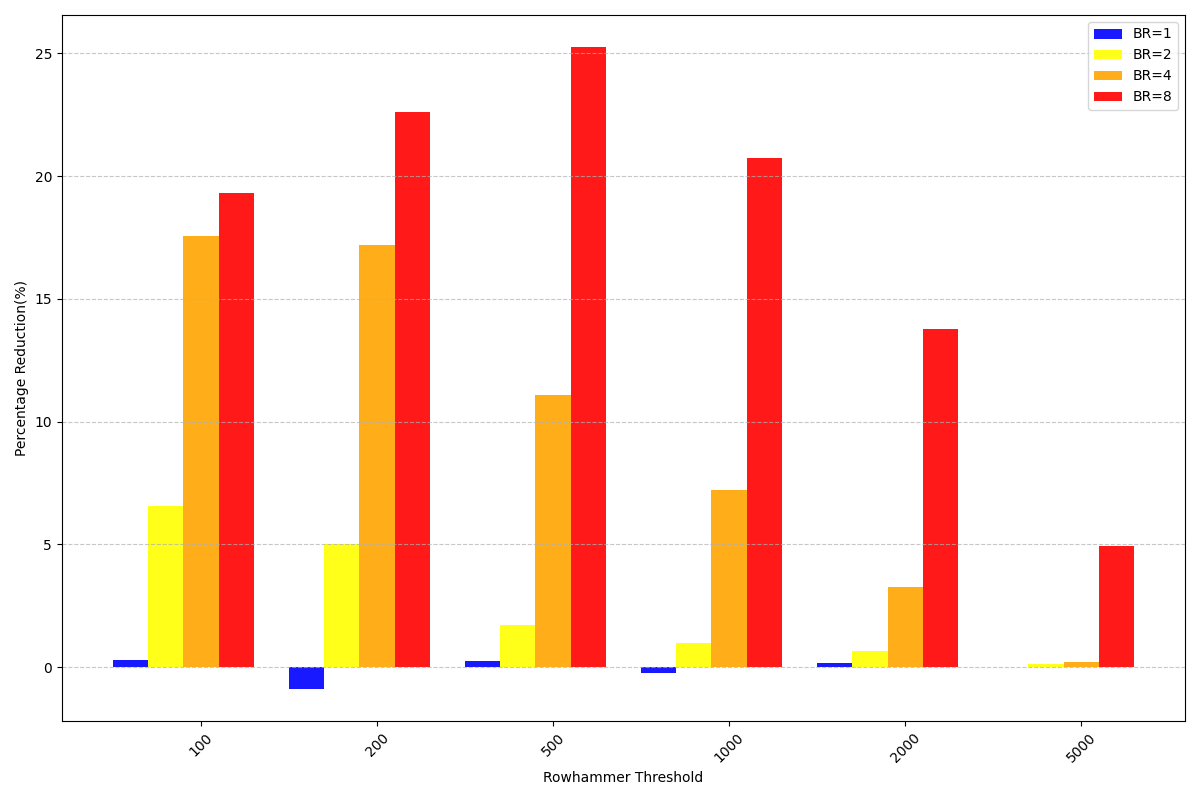}
\caption{Percentage reduction in IPC STP. Lower is better.}
\label{fig_stp}
\end{figure}
\begin{figure}[!t]
\centering
\includegraphics[width=3.5in]{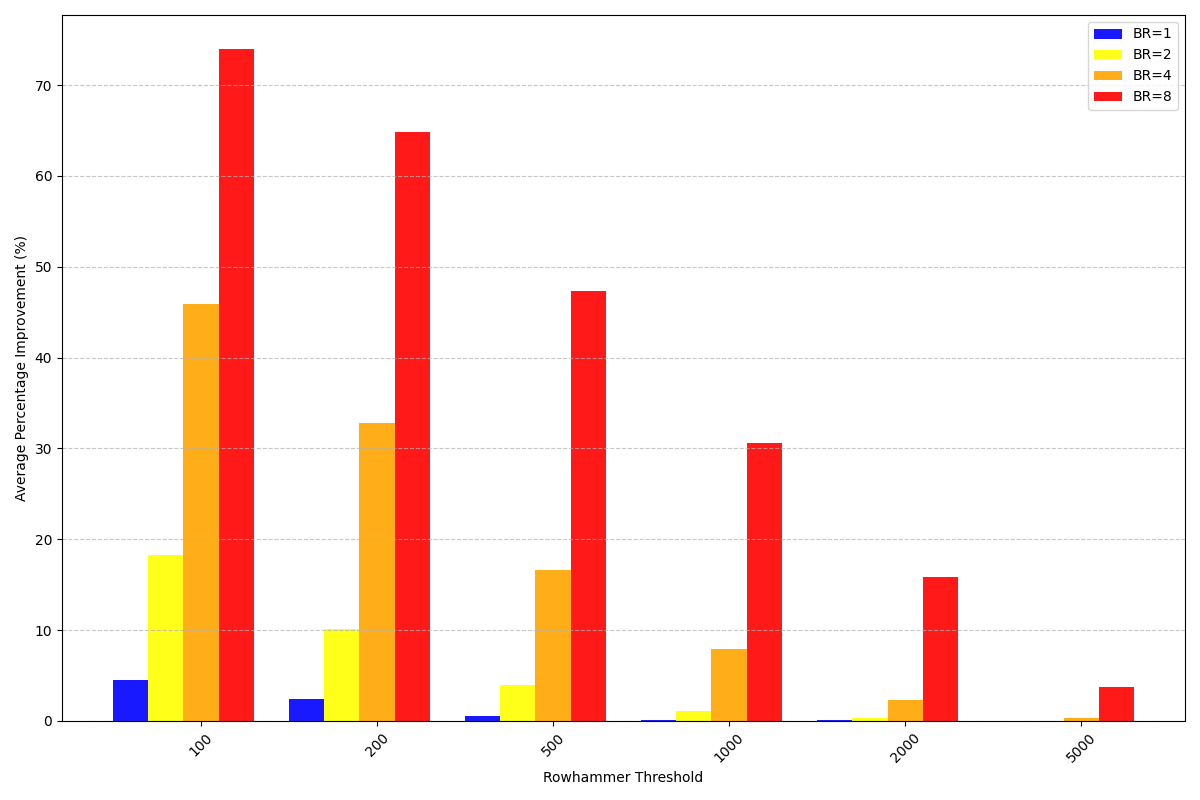}
\caption{Percentage improvement in Total Energy in DRAM}
\label{fig_tot_ener}
\end{figure}
\section{Conclusion}
In this work, we propose the Rowhammer Vulnerability Counter (RVC) technique, which shifts the focus from traditional aggressor-based activation count tracking methods to a victim-centric approach. We show that RVC consistently outperforms Graphene in detecting and preventing Rowhammer-induced bit-flips across varying blast radii. These findings highlight the potential of RVC as an effective mitigation strategy for modern DRAM systems. By maintaining a consistent threshold of \(T_{\text{th}}/2\) and selectively refreshing only the rows truly at risk, RVC minimizes unnecessary mitigations while optimizing system performance and energy efficiency. As technology scaling brings DRAM cells closer, leading to an increased blast radius, RVC's victim-focused approach becomes even more advantageous in ensuring reliable memory operation under these evolving conditions. This work thus introduces a novel and unexplored perspective on detecting Rowhammer, paving the way for more effective and efficient mitigation strategies.
\section{Future Work and Research Directions}
With the advent of DDR5 and its in-built Per Row Activation Counting (PRAC) capabilities, alternative strategies can be envisioned. For instance, it may be possible to embed the Rowhammer Vulnerability Count (RVC) directly within the DRAM itself, reducing overhead and latency. Additionally, novel methods to maintain and track the RVC, instead of relying on traditional aggressor activation counts, should be explored.
Finally, evaluating existing RowHammer detection techniques through the lens of RVC rather than aggressor counts—could offer new insights into improving efficiency and accuracy in RowHammer defense mechanisms.
\bibliographystyle{ieeetr}
\bibliography{references}

\vfill

\end{document}